\documentclass[runningheads]{llncs}
\usepackage[T1]{fontenc}
\usepackage{graphicx}
\begin{document}

\title{KneeXNeT: An Ensemble-Based Approach for Knee Radiographic Evaluation}

\titlerunning{An Ensemble-Based Approach for Knee Radiographic Evaluation}
%
\author{Nicharee Srikijkasemwat\inst{1}\thanks{Corresponding author: nicharee.srikijkasemwat@some.ox.ac.uk}\and
Soumya Snigdha Kundu\inst{2}\and
Fuping Wu\inst{3}\and
Bart\l omiej W. Papie\.z\inst{3}}
\authorrunning{N. Srikijkasemwat et al.}
%
\institute{Institute of Biomedical Engineering, Department of Engineering Science, University of Oxford, Oxford, UK \and
Department of Surgical \& Interventional Engineering, King's College London, London, UK \and
Nuffield Department of Population Health, Big Data Institute, University of Oxford, Oxford, UK}
\maketitle              
\begin{abstract}
Knee osteoarthritis (OA) is the most common joint disorder and a leading cause of disability. Diagnosing OA severity typically requires expert assessment of X-ray images and is commonly based on the Kellgren-Lawrence grading system, a time-intensive process. This study aimed to develop an automated deep learning model to classify knee OA severity, reducing the need for expert evaluation. First, we evaluated ten state-of-the-art deep learning models, achieving a top accuracy of 0.69 with individual models. To address class imbalance, we employed weighted sampling, improving accuracy to 0.70. We further applied Smooth-GradCAM++ to visualize decision-influencing regions, enhancing the explainability of the best-performing model. Finally, we developed ensemble models using majority voting and a shallow neural network. Our ensemble model, KneeXNet, achieved the highest accuracy of 0.72, demonstrating its potential as an automated tool for knee OA assessment.
\end{abstract}
\section{Introduction}
\label{sec:intro}
Osteoarthritis (OA) is a degenerative joint disorder and one of the leading causes of disability, particularly in the elderly population \cite{Intro1}. Among the joints affected by OA, the knee is the most commonly impacted, with a global prevalence of knee OA reaching 22.9\% among individuals aged 40 years and older \cite{Intro2,prevalence}. While age is a primary risk factor, other contributors to knee OA include gender, genetic predispositions, obesity, prior injuries, physical inactivity, and lifestyle factors \cite{Intro2,Intro1,strategy,OA_features}.
Patients with knee OA often suffer from knee pain, joint stiffness, swelling, and challenges in performing daily activities \cite{strategy,OA_features}. In advanced stages, the condition can lead to significant disability and reduced quality of life \cite{irreversible}.
Various imaging techniques can aid in diagnosing knee OA, but X-ray imaging is the most commonly used due to its low cost and widespread availability \cite{irreversible}. Key radiographic features of knee OA include joint space narrowing, osteophyte formation, cyst formation, subchondral sclerosis, and coronal tibiofemoral subluxation \cite{OA_features}.
The severity of knee OA is typically classified using the Kellgren-Lawrence (KL) grading system, which categorizes OA progression into five grades: None (Grade 0), Doubtful (Grade 1), Minimal (Grade 2), Moderate (Grade 3), and Severe (Grade 4) \cite{KLgrading}. Each stage of OA requires tailored treatment; for instance, exercise is recommended at early stages, while severe OA may necessitate joint replacement \cite{strategy,Intro1}. Early diagnosis and grading are essential to slow disease progression and guide treatment strategies, as untreated OA can advance to an irreversible stage \cite{strategy,irreversible}.
Although radiographic evaluation can be repeated frequently, the interpretation of these images requires expert radiologists and can be time-consuming. Additionally, the subtle changes associated with early OA make accurate grading challenging. In response to these needs, this project aims to develop a deep learning model capable of automatically classifying knee OA severity, thereby supporting clinicians in evaluating knee radiographs more efficiently.

Recently, deep learning techniques have been widely adopted for knee OA classification, with ensemble methods emerging as particularly effective due to their high performance \cite{Mikhaylichenko,ensemble_kneexray}. While previous studies have achieved promising results, our work aims to expand upon these by incorporating a broader range of deep learning models and investigating different ensemble strategies to enhance knee OA classification.

Contributions of our work can be summarised as follows. 
First, we evaluated the performance of ten state-of-the-art deep learning models on a publicly available knee OA X-ray dataset.
Secondly, to address the significant class imbalance in the dataset, we applied a weighted sampling strategy to improve model training.
To further enhance classification performance, we explored two ensemble methods—majority voting and a shallow neural network. These ensemble techniques demonstrated their effectiveness in OA grading, achieving an overall accuracy of 0.72, a notable improvement over individual models.

\section{Method}
\label{sec:Method}

\subsection{Dataset}
This study employed the Osteoarthritis Initiative (OAI) dataset, which categorizes knee OA severity into five classes based on the Kellgren-Lawrence (KL) grading system. The dataset comprises 8,260 knee X-ray images from 4,796 participants aged 45 to 79 years \cite{KneeOA_paper}.

We split the data into training, validation, and test sets in a 7:1:2 ratio, yielding 5,778, 826, and 1,656 images in each set, respectively. The same unseen test set was used to evaluate all models developed in this study. Each image was resized to 224 \(\times\) 224 pixels to standardize input dimensions across models.
As shown in Fig.~\ref{fig:Class_distribution}, the dataset exhibits significant class imbalance, which is a common challenge in medical image analysis.

\subsection{Baseline models}
In this project, we evaluated the performance of 10 state-of-the-art deep learning models for knee OA classification, including ResNet-18, ResNet-34, ResNet-50 \cite{Resnet}, VGG-16, VGG-19 \cite{VGG}, MobileNet \cite{Mobilenet}, DenseNet-121, DenseNet-161 \cite{Densenet}, EfficientNet \cite{Efficientnet}, and GoogLeNet \cite{Googlenet}. All models were pretrained on the ImageNet dataset \cite{russakovsky2015imagenet} prior to fine-tuning on the knee OA dataset.

Each model was trained for 30 epochs with a batch size of 28, using the Adam optimizer \cite{kingma2014adam} with an initial learning rate of 0.0001, reduced by a factor of 10 every 5 epochs. After training, each model was evaluated on the validation set, and the model weights corresponding to the highest validation accuracy were saved. To enhance data variety, we applied augmentation techniques, including random horizontal flipping (with a probability of 0.5), brightness adjustment (factor range: 0.5 to 1.2), saturation adjustment (factor range: 0.5 to 1.5), rotation (within 5 degrees), and random translation (within 10\% of the image size).

Two experiments were conducted to assess model performance for OA grading as follows.
Experiment~1: We compared the performance of 10 state-of-the-art models trained with cross-entropy loss to establish baseline accuracy.
Experiment~2: To address class imbalance, we introduced a weighted sampling strategy in training, assigning probabilities to each class inversely proportional to the class sample size. Each model was trained and tested three times, and we reported the mean and standard deviation of F1 scores and test accuracy

\subsection{Ensemble models}
To improve classification performance, we combined all 10 models from each experiment into an ensemble model using two strategies: majority voting and a shallow neural network.

In the majority voting strategy, each input image was passed through all 10 models, each outputting five logits corresponding to the classes (0 to 4). These logits were converted to class probabilities using the softmax function. The probabilities for each class across the 10 models were then summed, and the class with the highest combined probability was selected as the final prediction.

In the shallow neural network method, we designed a two-layer fully connected neural network (FCN) to perform the ensembling. For each input image, the logits from all 10 trained models were concatenated into a 50-dimensional vector, which served as the input to the FCN. The FCN was trained for 30 epochs using cross-entropy loss, with the same training parameters as the baseline models. To ensure consistency, training and testing were repeated three times, with the mean and standard deviation of results reported.

\begin{figure}
    \centering
    \includegraphics[width=1\linewidth]{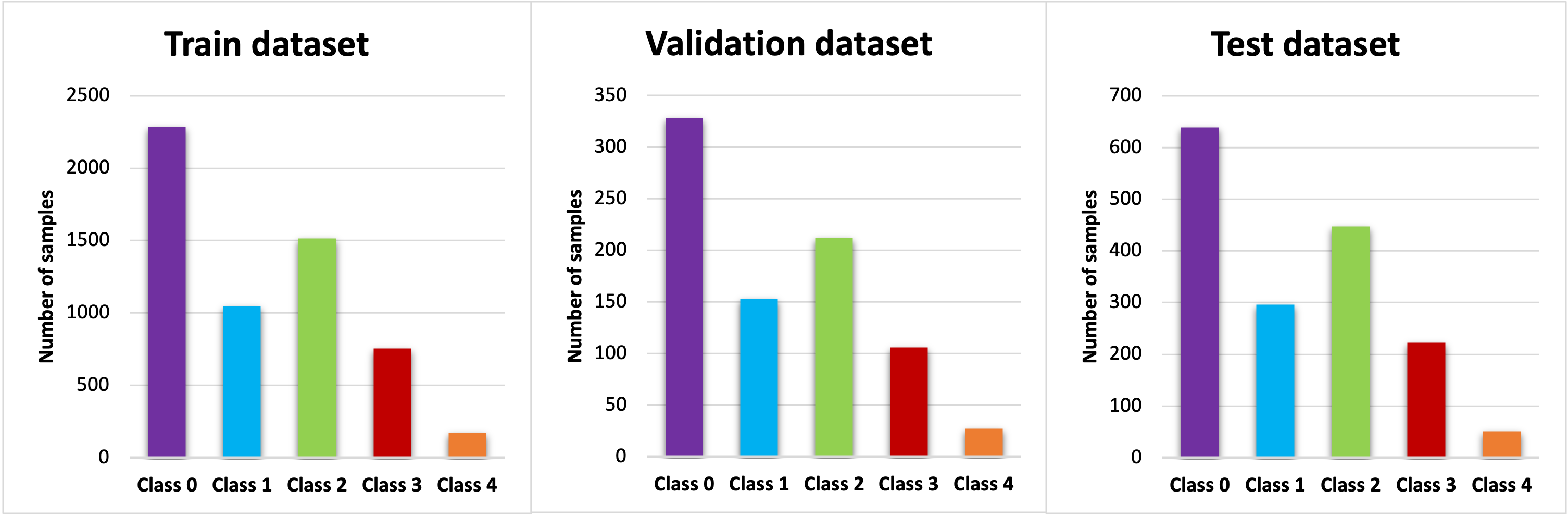}
    \caption{Distribution of samples across each class in the training, validation, and test sets.}
    \label{fig:Class_distribution}
\end{figure}

\section{Results}
\label{sec:results}

\subsection{Experiment 1: Comparison of the baseline models} 
In this experiment, we trained 10 models using cross-entropy loss to evaluate the performance of each model. As shown in Fig.~\ref{fig:Test_accuracy}, the highest accuracy of 0.69 was achieved by multiple networks ResNet-34, ResNet-50, VGG-16, VGG-19, DenseNet-121, and DenseNet-161.

Additionally, as observed in Fig.~\ref{fig:F1_score_exp1}, the F1 score for Class 0 (None) was generally higher compared to Classes 1 (Doubtful) and 2 (Minimal). This trend can be attributed to the class imbalance in the dataset, as shown in Fig.~\ref{fig:Class_distribution}. Notably, Classes 3 (Moderate) and 4 (Severe) exhibited the highest F1 scores across all models. This is likely due to the fact that severe OA conditions are easier to identify, despite the smaller number of images in these classes.

\subsection{Experiment 2: The baseline models with the weighted sampling strategy} 
\begin{figure}
    \centering
    \includegraphics[width=1.0\linewidth]{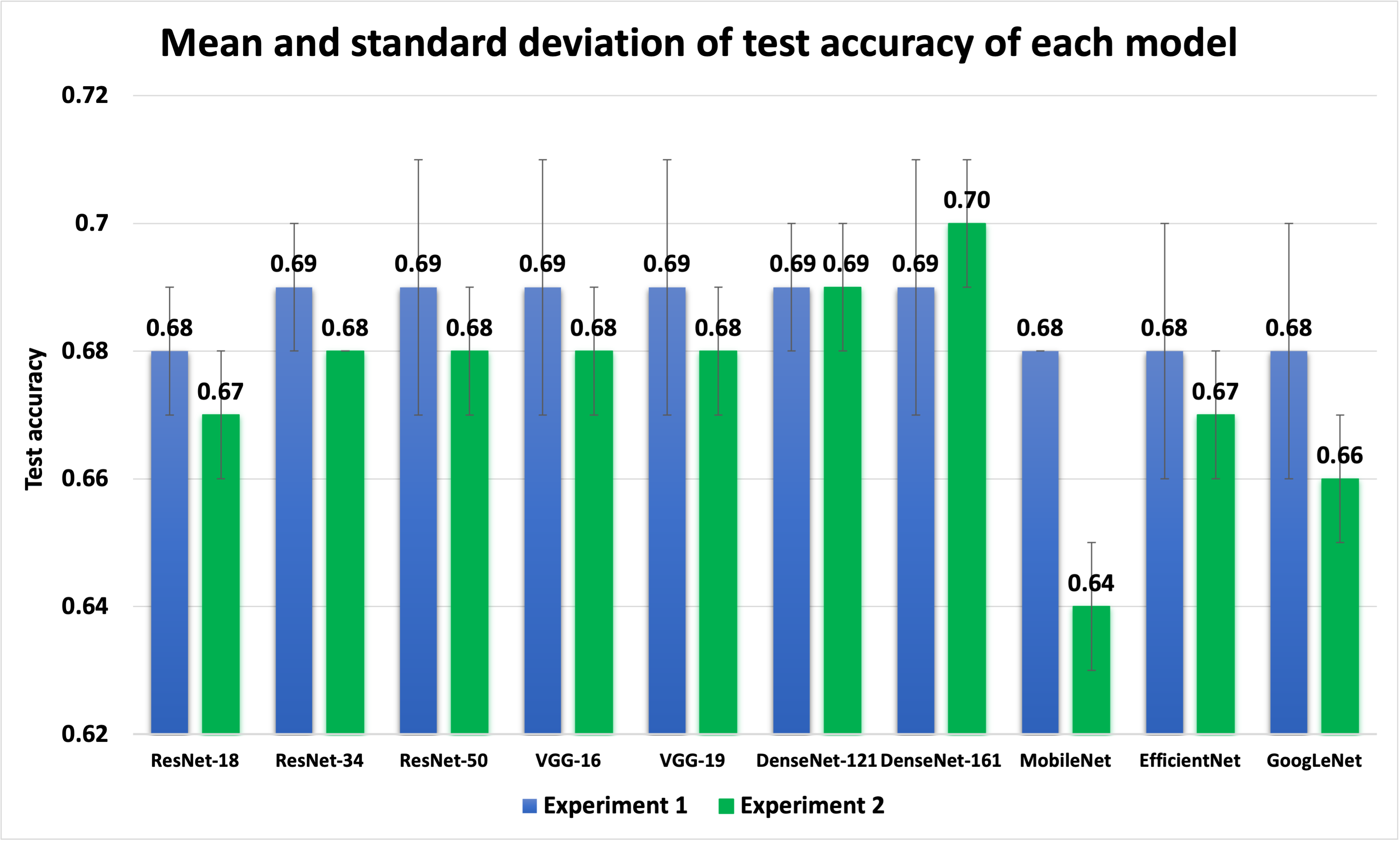}
    \caption{Test accuracy for each model in Experiment 1 and Experiment 2.}
    \label{fig:Test_accuracy}
\end{figure}

In Experiment 2, we investigated whether the weighted sampling method could improve model performance, particularly for Class 1, which exhibited low F1 scores in Experiment 1 due to class imbalance. As shown in Fig.~\ref{fig:Test_accuracy}, DenseNet-161 achieved the highest accuracy (0.70 ± 0.01) among all models, surpassing the best-performing model from Experiment 1. However, most models showed a slight decrease in accuracy compared to Experiment 1.

\begin{figure}
    \centering
    \includegraphics[width=1\linewidth]{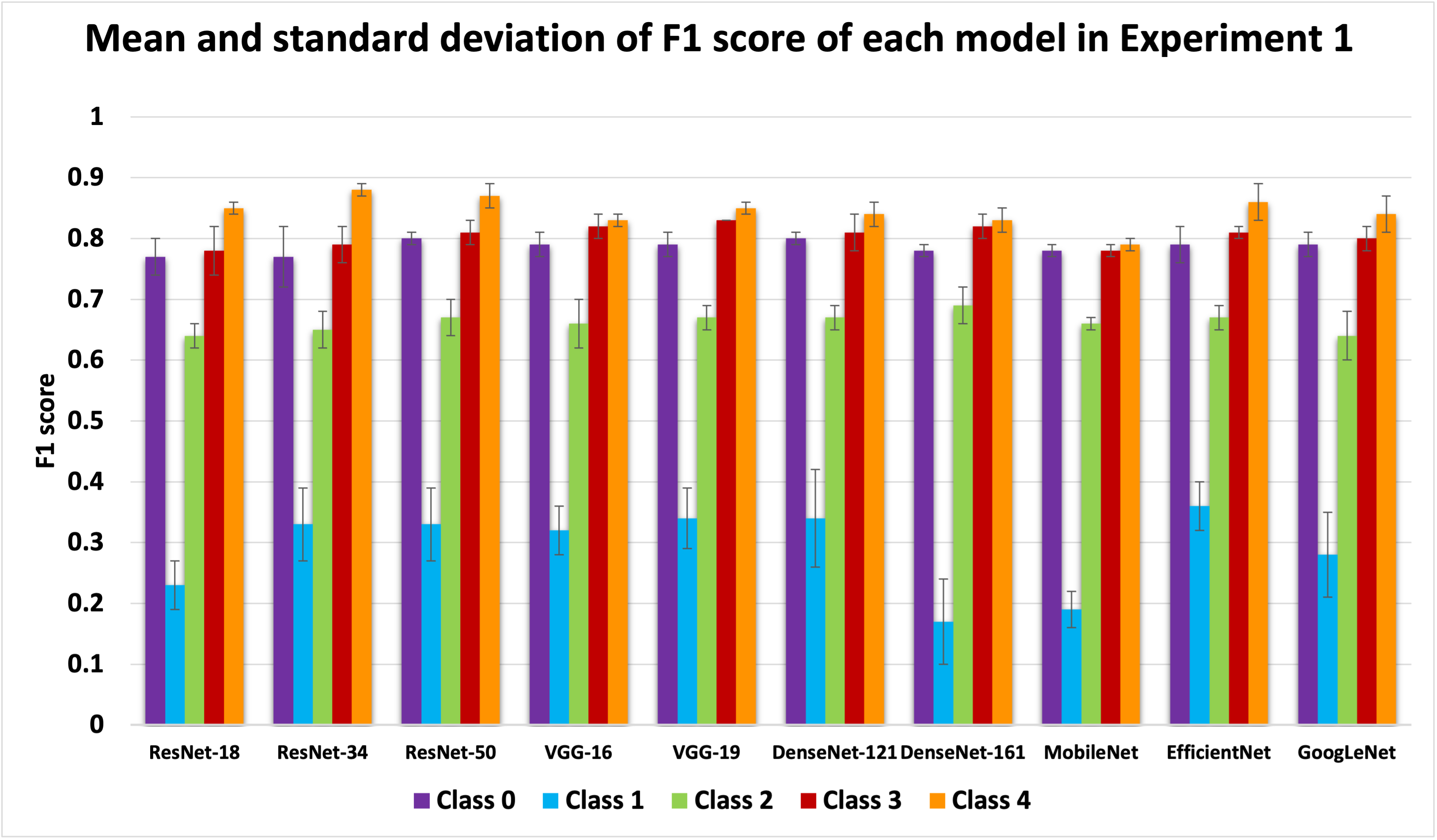}
    \caption{F1 scores of each model in Experiment 1 (the baseline models).}
    \label{fig:F1_score_exp1}
\end{figure}

\begin{figure}
    \centering
    \includegraphics[width=1\linewidth]{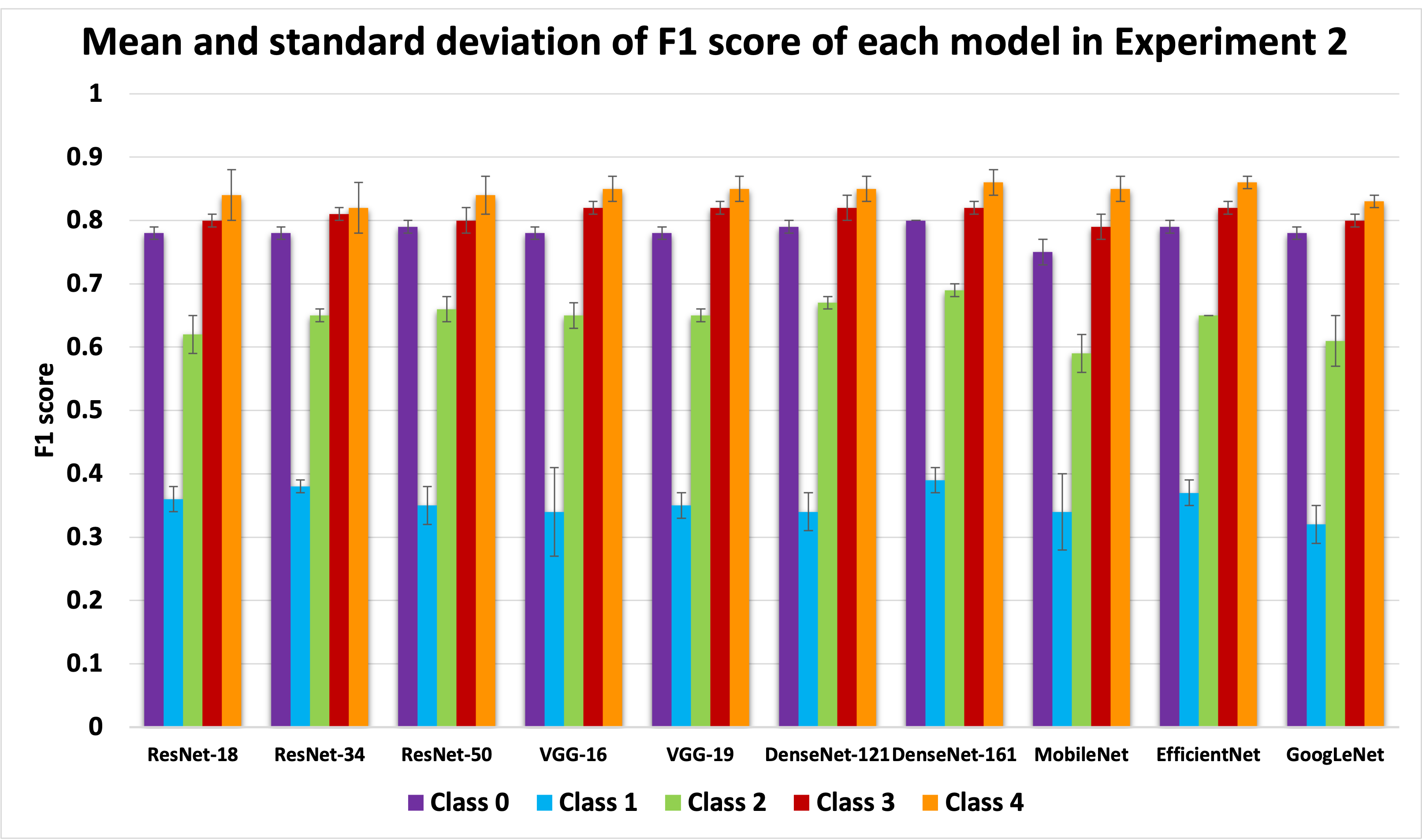}
    \caption{F1 scores of each model in Experiment 2 (the weighted sampling strategy).}
    \label{fig:F1_score_exp2}
\end{figure}

Additionally, the F1 scores for Class 1 in Experiment 2 (Fig.~\ref{fig:F1_score_exp2}) improved across all models compared to Experiment 1(Fig.~\ref{fig:F1_score_exp1}), indicating that the weighted sampling method successfully addressed the class imbalance. While F1 scores for some other classes decreased slightly, the overall improvement in Class 1 demonstrates the limited effectiveness of this method in mitigating the impact of class imbalance on model performance for the given data set.

\begin{figure}
    \centering
    \includegraphics[width=1.0\linewidth]{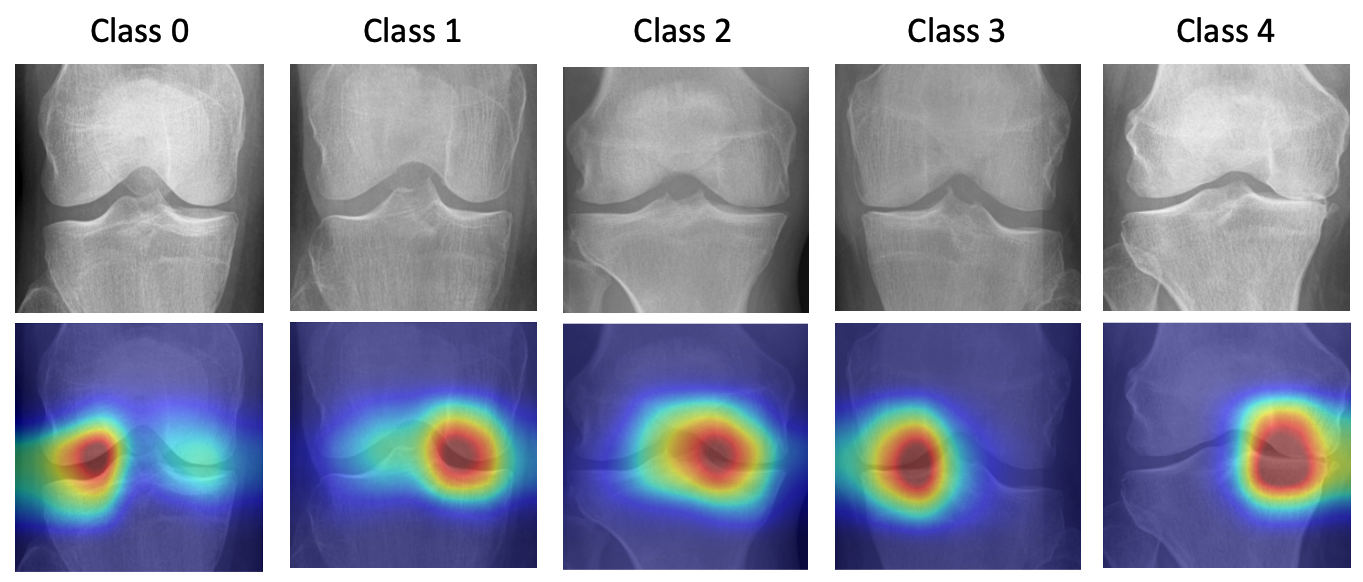}
    \caption{Visualizations of the regions influencing model predictions using Smooth Grad-CAM++. These heatmaps highlight the areas of the knee X-ray that the model focuses on when making predictions.}
    \label{fig:Explainability}
\end{figure}

\subsection{Ensemble model from Experiment 1}
To improve overall model performance, two ensemble techniques were applied and compared. In the first technique, the 10 trained models from Experiment 1 were combined using majority voting. The resulting ensemble model achieved a test accuracy of 0.70, which was slightly higher than the best-performing model from Experiment 1. The F1 scores for each class were 0.81, 0.21, 0.70, 0.80, and 0.76, respectively. Despite the improvement in overall accuracy, the F1 score for Class 1 (Doubtful) did not improve compared to Experiment 1. Since none of the individual models was particularly effective at identifying Class 1 images, the ensemble model did not show significant improvement in this class.

Next, the 10 models were ensembled using a shallow neural network. This method yielded a test accuracy of \(0.72\pm0.01\), which was higher than the result from majority voting.
The F1 scores for each class were \(0.82\pm0.01\), \(0.23\pm0.07\), \(0.71\pm0.01\), \(0.84\pm0.00\), and \(0.88\pm0.01\). The F1 scores for all classes, except Class 1, showed improvement compared to the individual models in Experiment~1.

\subsection{Ensemble model from Experiment 2}
The same ensemble strategies were applied to the 10 models trained in Experiment 2. Using majority voting, the second ensemble achieved a test accuracy of 0.64. The F1 scores for each class were 0.75, 0.33, 0.62, 0.76, and 0.82, respectively. This performance was lower than most of the individual models from Experiment 2.

Next, the outputs of the 10 models were fused using a shallow neural network. This ensemble achieved an accuracy of \(0.70\pm0.01\), which was higher than the majority voting technique and matched the performance of DenseNet-161 from Experiment~2. The F1 scores for each class were \(0.80\pm0.01\), \(0.37\pm0.03\), \(0.68\pm0.01\), \(0.84\pm0.01\), and \(0.88\pm0.01\). Notably, the F1 score for Class 1 in this ensemble model was 0.14 higher than the one in Experiment 1’s ensemble, though the overall accuracy was 0.02 lower.

\subsection{Model explainability using Smooth-GradCAM++}
To assess the explainability of the model's predictions, Smooth-GradCAM++ \cite{SmoothGradCAMpp} was applied to the best-performing model, DenseNet-161, which was trained using the weighted sampling strategy. DenseNet-161 achieved the highest accuracy among the single models and was selected to investigate the model's decision-making process.

As shown in Fig.~\ref{fig:Explainability}, the model's attention was focused on the region between the bones, potentially indicating the joint space narrowing, which is a key feature used to classify OA severity \cite{OA_features}. This visualization demonstrates how the model relies on clinically relevant features to make predictions.

\section{Discussion and Conclusion}
\label{sec:Conclusion}
In this paper, we presented an investigation into the performance of 10 state-of-the-art deep learning models for classifying knee osteoarthritis severity from X-ray images using the Kellgren-Lawrence grading system. The highest accuracy of 0.69 was achieved by ResNet-50, VGG-16, VGG-19, DenseNet-161, ResNet-34, and DenseNet-121, with ResNet-34 and DenseNet-121 being the most robust models, exhibiting also the smallest standard deviation (\(0.01\)).
However, the individual models were less effective at classifying Class 1 (Doubtful) images, which we hypotetised that could be attributed to class imbalance. To address this, we explored a weighted sampling strategy during training, which improved the F1 score for Class 1, although the overall accuracy did not show a significant increase. DenseNet-161, trained with this sampling strategy, achieved the highest accuracy of \(0.70\pm0.01\).

The lower performance of Class 1 compared to other classes may also be due to the subtle differences between Class 1 (Doubtful) and Classes 0 (Healthy) or 2 (Minimal) \cite{hart2003kellgren}. One potential improvement could involve merging Class 1 with either Class 0 or Class 2, thereby reducing the number of classes and possibly making the classification task easier for the model.
Additionally, our results suggest that using a classification approach to directly mimic the KL grading system might not be the most optimal strategy, as OA is a progressive disease with no clear boundaries between the different KL grades \cite{felson2011defining}.

Additionally, ensemble techniques, including majority voting and shallow neural networks, were explored to further improve performance. The best ensemble model, which fused the outputs of models trained with and without the weighted sampling strategy using a shallow neural network, achieved the highest accuracy of (\(0.72\pm0.01\)).
 This result outperformed a previous ensemble approach that combined three DenseNet-121 models trained with different random seeds, which achieved an average accuracy of 0.71 for multi-class classification \cite{Mikhaylichenko}.

This approach has the potential to assist clinicians in diagnosing knee OA more efficiently and accurately. It may prove particularly valuable in hospitals or clinics where access to specialists for radiograph interpretation is limited or unavailable.

\subsubsection{Acknowledgements} N.S. would like to thank Anissa Alloula for providing suggestions on using \textit{WeightedRandomSampler} method on this data to improve the model performance.

The computational aspects of this research were supported
by the Wellcome Trust Core Award Grant Number 203141
/Z/16/Z and the NIHR Oxford BRC. The views expressed are
those of the author(s) and not necessarily those of the NHS,
the NIHR or the Department of Health

The project was funded by Oxford Big Data Summer Internship Programme 2023.
\subsubsection{Compliance with ethical standard}
This research study was conducted retrospectively using human subject data made available in open access by Osteoarthritis Initiative (OAI) dataset \cite{dataset}. Ethical approval was not required as confirmed by the license attached with the open access data.

%
%
%
\bibliographystyle{splncs04}
\bibliography{mybibliography}
\end{document}